\documentclass{aa}

\usepackage[utf8]{inputenc}
\usepackage{graphicx}
\usepackage{natbib}
\bibpunct{(}{)}{;}{a}{}{,} 

\usepackage{txfonts}

\usepackage{xcolor}

\newcommand{\ds} {$\delta$ Scuti}
\newcommand{\corot} {CoRoT}
\newcommand{\kepler} {\emph{Kepler}}

\newcommand{\arma} {ARMA}
\newcommand{\stara} {HD~174966}
\newcommand{\starb} {HD~51193}
\newcommand{\starc} {HD~49933}
\newcommand{\cd } {$\mbox{cd}^{-1}$}

\begin{document}

  \title{MIARMA: An information preserving method for filling gaps in time series}
  \subtitle{Application to CoRoT\thanks{The CoRoT space mission, launched on December 27th 2006, has been developed and is operated by CNES, with the contribution of Austria, Belgium, Brazil, ESA (RSSD and Science Program), Germany, Spain.} light curves}

   \author{J. Pascual-Granado
          \inst{1}
          \and
          R. Garrido
          \inst{1}
          \and
          J. C. Suárez
          \inst{1}
          }

   \institute{Instituto de Astrofísica de Andalucía - CSIC, 18008, Granada, Spain\\
              \email{javier@iaa.es, garrido@iaa.es, jcsuarez@iaa.es}
             }

  \abstract
   {Gaps in time series originate spurious frequencies in the power spectra. In the case of light curves of pulsating stars, this hampers the identification of the theoretical oscillation modes. When these gaps are small as compared with the total time span, the usual approach to overcome these difficulties involves linear interpolation. In any case the resulting time series has not the property of preserving the original frequency content of the pulsating stars.}
   {The method here presented intends to minimize the effect of the gaps in the power spectra by gap-filling preserving the original information, that is, in the case of asteroseismology, the stellar oscillation frequency content.}
   {We make use of a forward-backward predictor based on autoregressive moving average modelling (\arma) in the time domain. The method MIARMA is particularly suitable for replacing invalid data such as those present in the light curves of the \corot\ satellite due to the pass through the South Atlantic Anomaly, and eventually for the data gathered by the NASA planet hunter \kepler. We select a sample of stars from the ultra-precise photometry collected by the asteroseismic camera on board the \corot\ satellite: the \ds\ star \stara, showing periodic variations of the same order as the \corot\ observational window, the Be star \starb, showing longer time variations, and the solar-like \starc, with rapid time variations.}
   {We showed that in some cases linear interpolations are less reliable to what was believed. In particular: the power spectrum of \stara\ is clearly aliased when this interpolation is used for filling the gaps; the light curve of \starb\ presents a much more aliased spectrum than expected for a low frequency harmonic signal; and finally, although the linear interpolation does not affect noticeably the power spectrum of the \corot\ light curve of the solar-like star \starc, the \arma\ interpolation showed rapid variations previously unidentified that \arma\ interprets as a signal. In any case, the \arma\ interpolation method provides a cleaner power spectrum, that is, less contaminated by spurious frequencies. In conclusion, MIARMA appears to be a  suitable method for filling gaps in the light curves of pulsating stars observed by \corot\ since the method preserves their frequency content, which is a necessary condition for asteroseismic studies.}
   {}

   \maketitle
%

\section{Introduction}

As originally pointed by \cite{DEE75}, gaps change the original frequency content of any sampled signal, because the observed power spectrum is the result of a convolution of the original signal with the observational window. When the gaps are small as compared with the total time span, techniques based on gap-filling have been commonly used \citep{FAH82,BRO90,FOS99,ROQ00,GAR14}. For larger gaps, the prewhitening techniques become unavoidable \citep{BRE93}. 
Most of the popular gap-filling algorithms do not guarantee the preservation of the original frequency content, which is crucial for asteroseismology to be reliable. The most widely used technique, as it is the case for \corot\ data, is the linear interpolation because of its simplicity.

Space missions like \corot\ \citep{BAG06} or \kepler\ \citep{GIL10} have observed a large sample of stars with two main objectives: first the detection of transits to look for planets orbiting other stars and second, to characterize the stars through asteroseismology. Asteroseismology makes possible the determination of global properties of the stars like the radius or age, and allows also to infer the internal structure and internal rotation profile. For this objective a long and uninterrupted observation is required to obtain reliable results. In order to detect transits more precise measurements and a better time sampling is required. The requisites for both these objectives are fulfilled by the space missions, which are observing continuously and with an unprecedented resolution. Nevertheless, a photometric time series without gaps is an ideal case which is never reached. In practice, there are always some invalid flux measurements due to operational procedures like the change of mask, reorientation, data downloading, or environmental effects like the impact of energetic particles, as it is the case when \corot\ pass through the South Atlantic Anomaly (SAA). These produce invalid data or interruptions in the time series that affect the analyses in the frequency domain. 

We introduce here a new gap-filling method (MIARMA) which is based on a non-closed form and can represent any kind of function, even non-analytic functions, thus preserving the original frequency content of the signal. This method make use of a forward-backward predictor based on autoregressive moving average (\arma) models in order to fill the gaps present in astronomical time series. By applying this method to time series of stars with different pulsational characteristics observed by \corot\ satellite, new properties of the light curves are revealed.

We first review in Sect.2 some of the issues found when analysing gapped time series, which are reduced when appropriate interpolation methods are used, and we show how some of the usual interpolation methods are not appropriate, proving that an adequate gap-filling method could improve significantly the frequency determination of the oscillations in time series. In Sect.3 we describe the method for gap-filling based on modelling valid data segments as \arma\ processes. Then, in Sect.4 the application of the method is shown in three different cases: the \ds\ star \stara, the Be star \starb, and the solar-like \starc. A discussion follows in Sect.5 on the consequences of the results presented and the conclusions are presented in Sect.6.


\section{Gaps in time series}
\begin{figure}
   \centering
   \resizebox{\hsize}{!}{\includegraphics{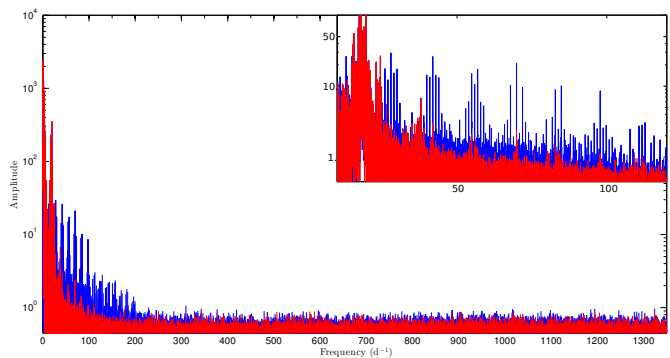}}
   \caption{Periodogram of the \ds\ star HD~172189 calculated using a Fast Fourier Transform of the linearly interpolated time series (blue), and as interpolated by MIARMA (red). Notice on the inset how the frequencies in the \ds\ instability range are affected by the spectral window.}
   \label{alias}
\end{figure}

In order to consider the effects of the gaps (meaning not only the lack of measurements but also invalid flux measurements) in the estimation of the power spectra one must consider the mathematical result due to \cite{DEE75}:
\begin{equation}
\label{eq-dee}
\frac{1}{N} F_N(\nu) = F(\nu)\ast W_N(\nu)
\end{equation}
That is, the Discrete Fourier Transform (DFT) of a gapped time series is the convolution of the Fourier Transform of the original function $F(\nu)$ with the spectral window $W_N(\nu)$, which is defined as the DFT of the function describing the observational time. 
All of the issues related with gaps in time series derive from this simple formula. First of all, the spectral window always contains a sinc function related to the length of the observation making the amplitudes of the peaks originated from harmonic components reduced because the power is spread in sidelobes. For long observations, as it is the case with space satellites, this effect is insignificant, but when the gaps are regularly distributed, another sinc function appears in the spectral window related to the length of the gaps that spreads much more power of the central peak in sidelobes - spectral leakage. As a consequence the signal-to-noise ratio is reduced leading to a less reliable frequency detection that affects parameter estimation.

The spectral window also can cause that spurious frequencies appear far from the central peak which can be confused with other peaks misleading the frequency detection and identification. 

On the other hand, $W_N(\nu)$ is a complex function so it introduces a deviation in the original phases of the signal. This deviation is difficult to determine when the gaps are not regularly distributed. This makes difficult the determination of the phases, which is crucial in the study of resonances in high amplitude pulsating stars \citep{GAR96} or in the determination of phase differences between different wavelengths for mode identification \citep{GAR00}.

Another issue concerns the normalisation of the periodogram to give a consistent estimation of the power spectrum. This problem was solved by \cite{SCA82} with a proper normalisation of the real and imaginary part that preserve the statistical properties of the regular sampling when the sampling is irregular. However, the Lomb-Scargle periodogram gives the same result as filling with zeros the gaps to obtain a regular sampling and then calculating the periodogram through a FFT. It can be shown that the convolution of the data with the spectral window introduce correlations between Fourier frequencies \citep{STA08}. In this way, as the Lomb-Scargle periodogram does not remove the effect of the spectral window, correlations between frequencies due to the gaps are maintained.
 
These effects have a lower impact in the case of \corot\ and \kepler\ than in ground-based observations since the duty cycle is always higher than $80\%$ and the gaps are not so regularly distributed as the diurnal cycle. However, it is still crucial to reduce them in order to obtain a consistent estimation of the power spectra. This is specially relevant in the case of \corot\ observations as the invalid data measurements introduced by the SAA are quasiperiodic \citep{AUV09}.
   
In order to obtain the power spectrum of a gapped time series many approaches have been used in the past. The most simple one is filling the gaps with zeros and then computing the periodogram of the evenly sampled time series with a FFT. This technique has the same pathologies that are found when using unequally-spaced data. Replacing the gaps with zeros does not avoid the spectral window contribution at all.

A second approach more commonly used is a simple linear interpolation between valid segments. This might be a good approach when the gaps and the variance of the time series are small, but in other cases this solution is insufficient to remove the aliases as can be seen in Fig.\ref{alias}. This figure shows the periodogram of the \corot\ level 2 data for the binary star HD~172189 which presents \ds\ pulsations. The gaps in the time series used to calculate the periodogram were filled by linear interpolation (details in Sect.~4). Note that when interpolating linearly the gaps the spurious peaks due to the spectral window appears clearly, but they are almost removed when the \arma\ interpolation method that we propose in the next section is used.

Other more sophisticated gap-filling techniques (e.g. \citeauthor{FOS99} 1999) used an approach based on the autocorrelation of the oscillation signals. 

More recently, the inpainting technique, based on a sparse representation of a dictionary of wavelet functions, is being used for filling gaps in \corot\ \citep{SAT10,MAT10,MAT13} and \kepler\ data \citep{GAR14}. However, all these methods for filling gaps make use of analytic functions and  cannot guarantee that the original frequency content is preserved in the filling procedure. Instead, we used a forward-backward predictor for gap-filling which takes only minimal data prior and after the gaps and interpolates using a non-closed form expression, that is, an autoregressive moving average process. This method do not require any representation a priori and allow analytic and non-analytic functions to be fitted.


\section{MIARMA: an \arma\ interpolation method}
\subsection{Autoregressive methods}
The aim of these methods is to determine a parametric representation of a time series assuming that the correlation between data can be expressed in a recursive formula like this:   
\begin{equation}
\label{eq-ar}
 x_n = a_0 + \sum_{k=1}^p a_k x_{n-k} + \epsilon_n 
\end{equation}
where $x_n$ is the time series, $a_k$ are the $p$ parameters of the autoregressive model AR(p) of order p, $\epsilon_n$ is the error and usually $a_0=0$ when the time series has zero mean. This representation can be used to model any kind of variability, analytic or not. For example, \cite{YUL27} showed that the solution of the differential equation of the damped harmonic oscillator can be expressed as a AR(2) process, that is:
\begin{equation*}
x_n = a_1 x_{n-1} + a_2 x_{n-2} + \epsilon_n.
\end{equation*}
Then, for time series originated from multiperiodic pulsating stars with M harmonic components it is reasonable to assume an AR representation with 2*M coefficients.

A moving average process (MA) is defined as:
\begin{equation}
\label{eq-ma}
 x_n = \sum_{k=1}^q b_k \epsilon_{n-k} + b_0 \epsilon_n
\end{equation}
where $\epsilon_n$ is an independent white noise process, $b_k$ are the q parameters of the model MA(q), and $b_0$ is usually normalised to 1. The moving average model can be identified with a linear filter with coefficients $b_k$ and input $\epsilon_n$ giving $x_n$ as the output.

These two representations are directly related: the MA representation is based on correlations between the inputs $\epsilon_n$ while the AR representation is based on correlations between the outputs $x_n$. Indeed, it can be demonstrated that an AR representation can be determined from a MA and vice versa.

On the other side, a mixed autoregressive moving average process \arma(p,q) can be expressed as:
\begin{equation}
\label{eq-arma}
 x_n = \sum_{k=1}^p a_k x_{n-k} + \sum_{k=1}^q b_k \epsilon_{n-k} + \epsilon_n.
\end{equation}
Notice that with the notation used above the AR(p) process represented in Eq.~\ref{eq-ar} can be also ARMA(p,0), and the MA(q) process represented in Eq.~\ref{eq-ma} can be also ARMA(0,q). Thus the \arma\ representation is a generalisation of AR and MA. Furthermore, some processes can be represented with a finite number of parameters in the ARMA representation while they require infinite parameters in a pure AR or MA. In those cases, an \arma\ model can give an exact representation of a signal while the others cannot and it is computationally more efficient. For a comprehensive discussion of these methods see \cite{SCA81}.

Likewise, we can state that the \arma\ representation is as general as Wold's theorem \citep{WOL38} permits, namely that any stationary signal can be represented as the sum of a deterministic (AR) plus a stochastic process (MA). In this way, these models are optimal since an exact representation of the signal can be found independently of its properties: periodic or aperiodic, deterministic or stochastic, analytic or non-analytic, chaotic or not.

\subsection{The algorithm}
Autoregressive methods have been used in the past in the context of pulsating stars for filling gaps. \cite{FAH82} used them for the first time with pure AR processes and global modelling. However, it can be easily seen that this method fails when the time series is non-stationary since the coefficients change during the observation. In \cite{ROQ00} an AR local model was used. This allows to reproduce seasonal changes in the time series and a certain non-stationarity is permitted, but the order is fixed by an arbitrary criteria, i.e. the average length of the data segments.    

\cite{ROT10} used a second order AR process whose parameters are determined through the Expectation Maximization algorithm. In that case, the amplitudes of the estimates depends on the duration of the time series and the computational cost is high because they have to model the whole time series globally too.

As far as we know these have been the only non-analytic approaches for filling gaps used in the field of asteroseismology.

Besides all the drawbacks mentioned, as only \arma\ provides an exact solution it can be considered as the most reliable method.

The gap-filling method we introduce here is similar to the original method of \cite{FAH82} but instead of AR models we used \arma\ models. Also, instead of a global model for the whole data set MIARMA implements local models. The algorithm basically consists in three steps: first, a criterion based on physical principles for selecting the optimal order of the \arma\ model; secondly, the data segments before and after a given gap are fitted using an \arma\ model of the order selected in the first step; next, the gap is interpolated using a weighted function of a forward and a backward prediction based on the models of the selected data segments. The second and third steps are repeated for each gap contained in the entire time series.

In the next sections we will describe throughly each of these steps.

\subsubsection{Criterion for selecting the order}
The criterion for selecting the order is a crucial step in this process (i.e. the number of harmonic components that can be fitted, that is, the frequency content, depends on the number of coefficients of the model). In this sense: too low values lead to models with insufficient spectral resolution, too high values lead to overfitting and inestability. We follow a statistical treatment for selecting the order based on the Akaike Information Criterion \citep{AKA74}.
This criterion is based on the maximization of the likelihood of the model parameters, that is, the quadratic mean error of the prediction \citep{ULR79}. The AIC parameter is defined as:
\begin{equation}
AIC = N\cdot\log V + 2\cdot d
\end{equation}
where V is the sum of the squares of the residuals for a given model, d is the number of parameters (for ARMA $d=p+q$) and N the number of data to be modelled. The quantity V is obtained by fitting an ARMA(p,q) model each time.
The first term of AIC formula quantifies the entropy rate or prediction error of the model. The second term penalizes the number of free parameters used by the model. According to Akaike's theory the optimal model is the one with lowest AIC value. The Akaike criterion derives from purely physical considerations, indeed it was called originally the Principle of Maximum Entropy. Therefore, this criterion is objectively self-consistent because it guarantees that the model found is the best approximation to the physical process observed using the available information.  
The procedure used here for selecting the order iterates from a set of models with different orders (p,q) and select the one with the minimum AIC value. In this way the optimal model for the data is always guaranteed whatever the range of the orders (p,q) explored. Note that in order to obtain a robust estimation of the order (i.e. valid for the entire time series) the longest non-interrupted data segment is selected for the calculations of the AIC coefficients.

\subsubsection{gap-filling}
Once the orders are selected using AIC, an \arma\ model is locally fitted to the data segment located before and after every gap. In order to do this the coefficients $a_k$ and $b_k$ appearing in Eq.\ref{eq-arma} have to be determined for each data segment. In the case of an AR model the parameters can be determined using minimum least squares, the moment method, or MCMC methods, but for an \arma\ model it is necessary to use an iterative procedure to obtain the coefficients of the MA part. The algorithm used by MIARMA is based on an iterative algorithm that minimizes the sum of the quadratic errors checking the convergence after a number of iterations.

The \arma\ interpolation in the gaps is based on the following equations: 
\begin{equation*}
x_n = x^f_n w_n^f + x^b_n w^b_n
\end{equation*}
\begin{equation*}
w^f_n = 1 - w^b_n = w.k, \qquad k \in (1, l_g)
\end{equation*}
\begin{equation}
w = \frac{1}{1 + l_g}
\end{equation}
where $x^f_n$ and $x^b_n$ are forward and backward predictions inside the gaps, $w^f_n$ and $w^b_n$ the weights normalizing those predictions respectively, and $l_g$ the length of the gap.

Usual algorithms for fitting \arma\ models are insensitive to the arrow of time, that is, they cannot make backward predictions. We solve this problem by mirroring the data segment after the gap, performing a forward prediction and subsequently mirroring the result again. Although in \cite{ROQ00} a backward prediction is used, as far as we know, our technique is a novel solution in the field. 

The whole gap-filling algorithm here described is carried out through each gap contained in the time series analyzed and the procedure is iterated until no gap is present or a predefined limit is reached.

Predictions lose coherence rapidly when the gaps are much more larger than the data segments modelled. In this way, the method is self-consistent because it can decides whether a prediction is no longer feasible attending the coherence loss. Unlike usual gap-filling techniques, which are based on analytic methods assuming that the pattern modelled in the data segments is stable and repeats until infinite, the method introduced here take into account the natural limits of prediction. It is possible to decide when a given prediction is not reliable (e.g. a possible result when applying this algorithm is that no interpolation is reliable). That is, contrary to analytic methods this algorithm guarantees that the predictions are reliable. In this sense MIARMA is optimised to preserve the frequency content of the time series, i.e., it guarantees that the conditions for Parseval's theorem to be valid are fulfilled.



\section{Results: CoRoT data}
The passing through the SAA introduce most of the invalid data (gaps) that are present in \corot\ observations \citep{AUV09}. As described in \cite{SAM07}, two sets are available in \corot\ level 2 data: the gapped data, and a regularly sampled dataset obtained by using linear interpolation in order to patch the invalid data produced during the SAA crossing.


Here we compare now the regularly sampled light curves obtained by using MIARMA with the original \corot\ level 2 data (linearly interpolated) in three cases of stars showing different pulsational characteristics: the \ds\ \stara, showing periodic variations of the same order as the \corot\ observational window, the Be star \starb, showing longer time variations, and the solar-like \starc, with rapid time variations. The data we use for the next sections were gathered by the ultrahigh precision CCD cameras onboard \corot\ satellite with the primary objective of studying stellar pulsations (seismofield camera).

\subsection{\stara}
The case of A-F main sequence stars is particularly critical because they show pulsation frequencies close to the orbital frequency of the satellite. In particular the \ds\ star \stara\ could be consider as a prototype to investigate the impact in this kind of variable stars.

\begin{figure}	
    \centering
    \resizebox{\hsize}{!}{\includegraphics{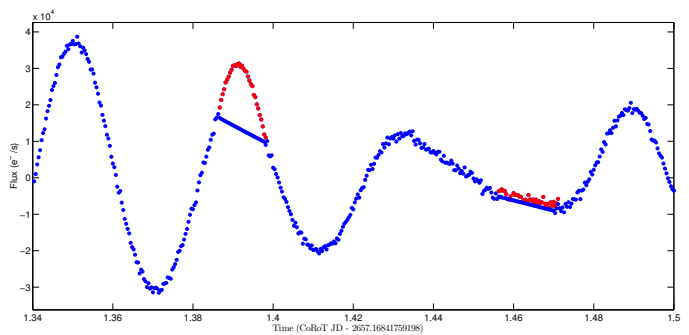}}
   \caption{Example of comparison between \arma\ gap-filling (red) and the linear interpolation (blue) for two gaps in the light curve of the \ds\ star \stara\ observed by \corot.}    
   \label{stara-lc}
\end{figure}

For this star it is clear that the linear interpolation does not preserve the signal (see Fig.~\ref{stara-lc}) in contrast to the \arma\ interpolation.

\begin{figure*}
   \centering
   \resizebox{\hsize}{!}{\includegraphics{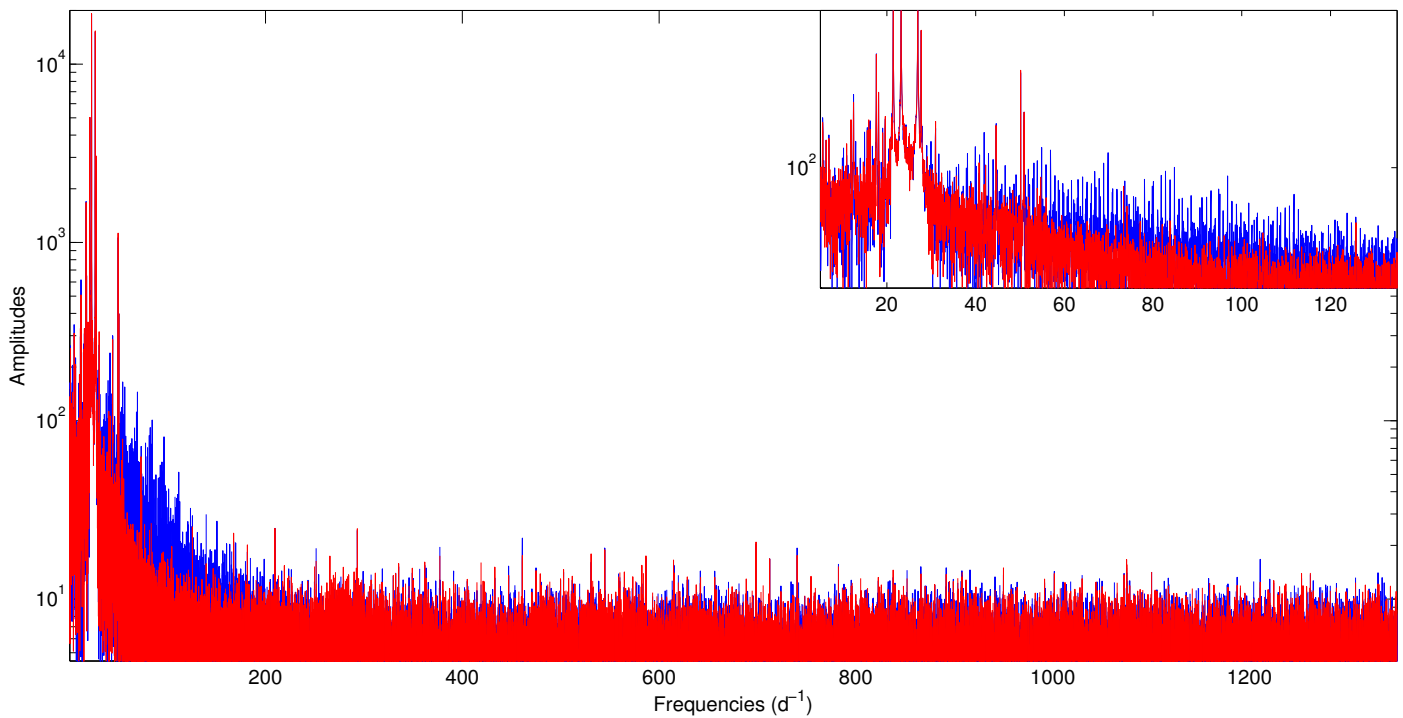}}
   \caption{Periodogram of the light curve of \stara\ observed by \corot\ after the \arma\ interpolation (red), and after the linear interpolation (blue). In the inset the range where \ds\ pulsations are excited is shown.}
   \label{stara-per}
\end{figure*}

The power spectrum of the linearly interpolated time series is seriously affected by the spectral window of \corot\ (see Fig.~\ref{stara-per}) in contrast with the \arma\ interpolation where the window is eliminated. The impact of the linear interpolation is so strong as the gapped data.

Definitively, the linear interpolation does not improve the original power spectrum biased by orbital aliases in this kind of pulsating stars.

\subsection{\starb}
\begin{figure*}	
    \centering
    \resizebox{\hsize}{!}{\includegraphics{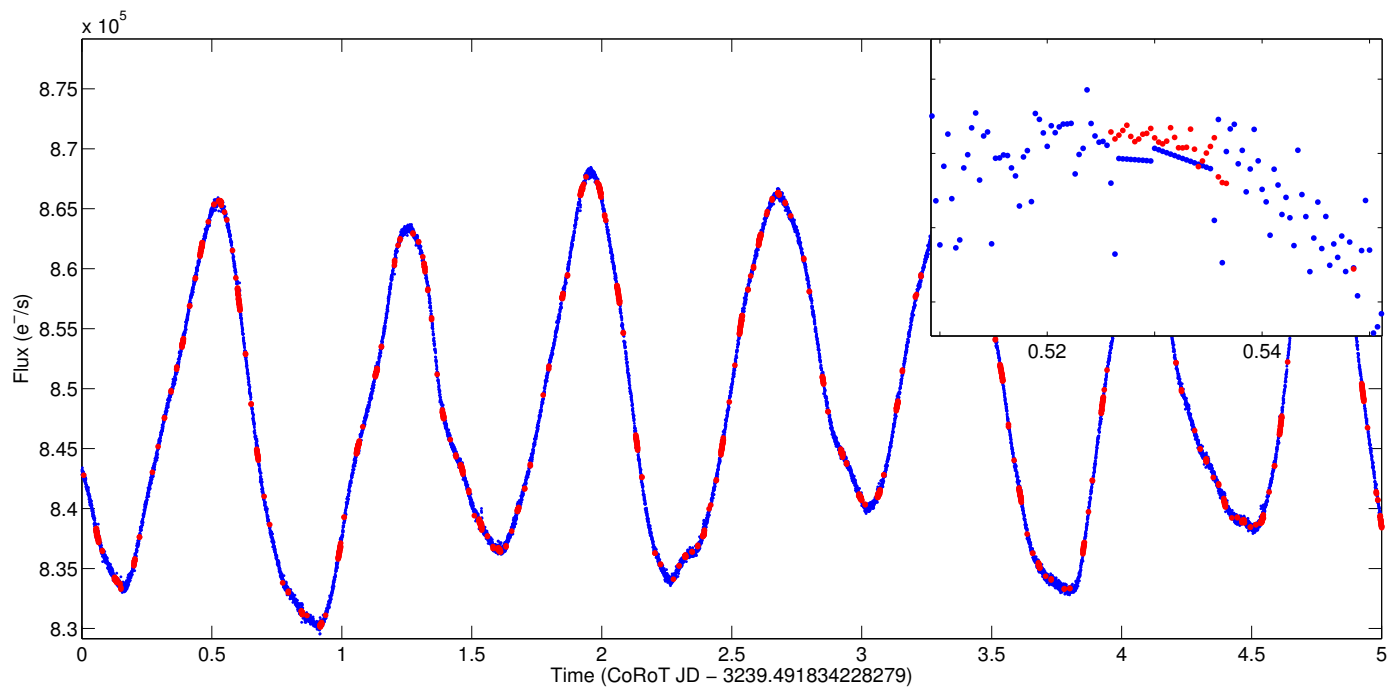}}
   \caption{Comparison between \arma\ gap-filling (red) and the linear interpolation (blue) for the light curve of the Be star \starb\ observed by \corot. In the inset a short segment of 1.44 hrs of duration is shown. Note that the size of gaps is around $\sim$0.014 ${\rm d}$, which one order of magnitude smaller than the pulsation period of the star (several days)}    
   \label{starb-lc}
\end{figure*}

\begin{figure*}
   \centering
   \resizebox{\hsize}{!}{\includegraphics{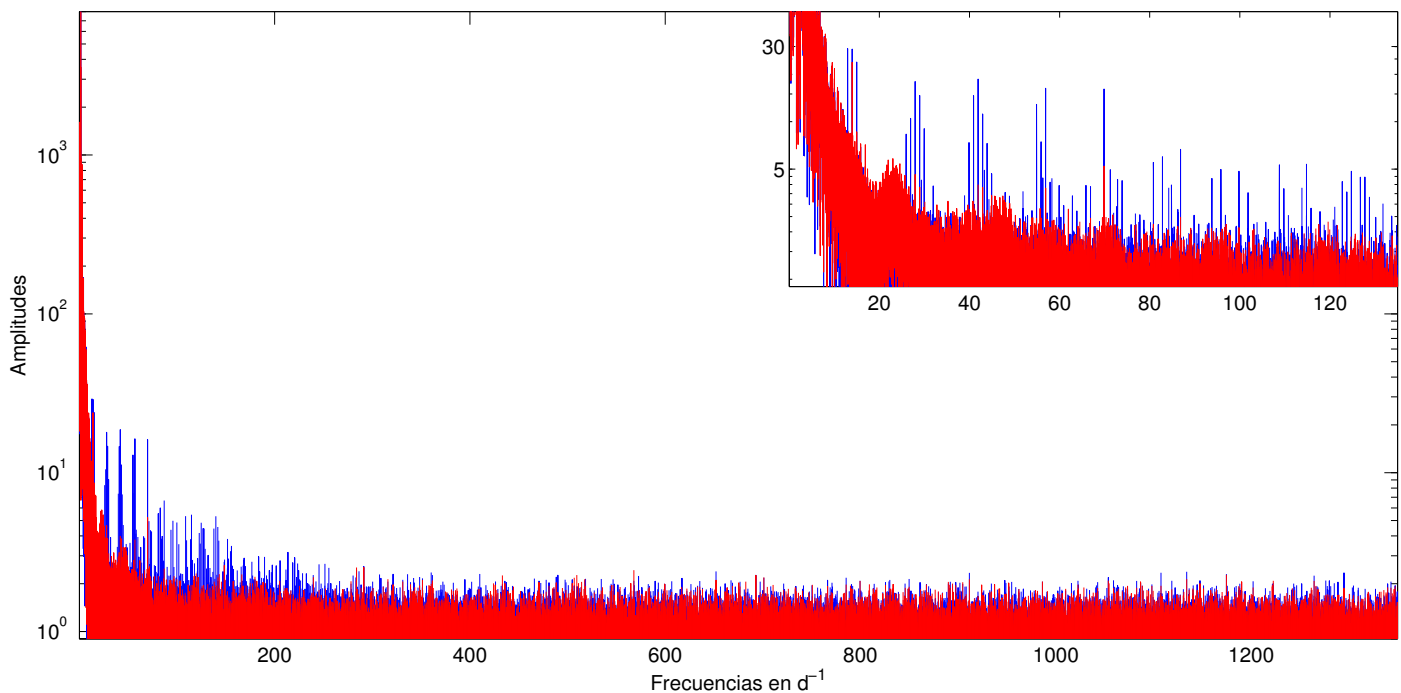}}
   \caption{Periodogram of \starb\ obtained after \arma\ interpolation (red) and after linear interpolation (blue). In the inset the range from 0 to 130 d$^{-1}$ is shown.}
   \label{starb-per}
\end{figure*}

When the period of variation is much larger than the size of the gaps, which is the case of Be stars, the filling provided by linear and \arma\ interpolation
looks similar. However, this is only a matter of scale as can be seen in the inset of Fig.~\ref{starb-lc} for the Be star \starb. It is commonly assumed that
so small gaps in the light curve have no effect on the power spectrum, however, we show here that for \starb\ the spectrum obtained from the linear interpolation provided by
the \corot\ pipeline is clearly biased by the spectral window of the orbital frequency (13.97 \cd), in contrast to the \arma\ interpolation, in which such 
an effect has been removed (see Fig.~\ref{starb-per}).

Due to the ultrahigh sensitivity of the instrument detectors, even the very small temperature variations within the technical specifications, are sufficient to make the instrument response a function of the orbit. This produces a modulation of the signal which is more significant for the frequencies with the highest amplitudes. Such a modulation appears in both spectra as a peak at 13.97 \cd, which is the orbital period of the satellite (inset of Fig.~\ref{starb-per}).

\subsection{\starc}

The solar-like star \starc\ was chosen to test MIARMA algorithm with star showing rapid variations, i.e. the CoRoT gaps contain several stellar pulsation cycles.
The optimal model for the longest segment without invalid data (640 datapoints of a total length of 369601 datapoints) is ARMA(37,0), i.e. a purely autoregressive model with 37 terms. 

\begin{figure*}	
   \centering
   \resizebox{\hsize}{!}{\includegraphics{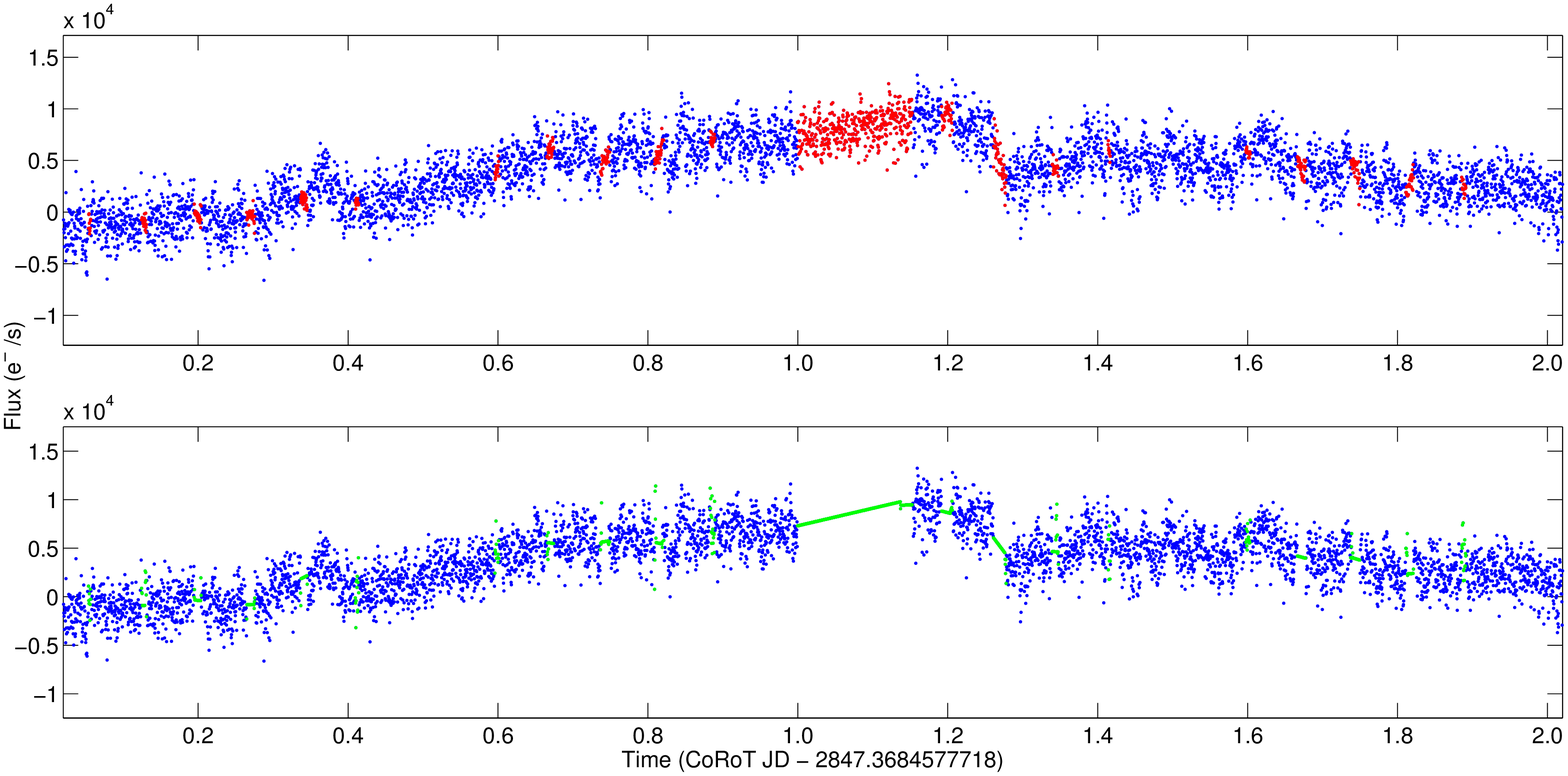}}
   \caption{Comparison between \arma\ interpolation (upper panel in red) and linear interpolation (lower panel in green) for the light curve of \starc\ observed by the \corot\ satellite. Valid data is depicted in blue (both panels).}    
   \label{starc-lc1}
\end{figure*}

 Although, the most frequent gap length is 9 minutes \citep{APP08}, some gaps last up to 0.2 days. Furthermore, this time series has a very high level of noise. All these make the linear interpolation quite erratic (see Fig.~\ref{starc-lc1}). However, in the interpolated segment between 1.0 and 1.2 days,  \arma\ predicts a signal with a kind of fine structure. We do not know the origin of such an structure, but we be confident (see next section) that  ARMA prediction does not correspond to a white noise process. 

\begin{figure*}
   \centering
   \resizebox{\hsize}{!}{\includegraphics{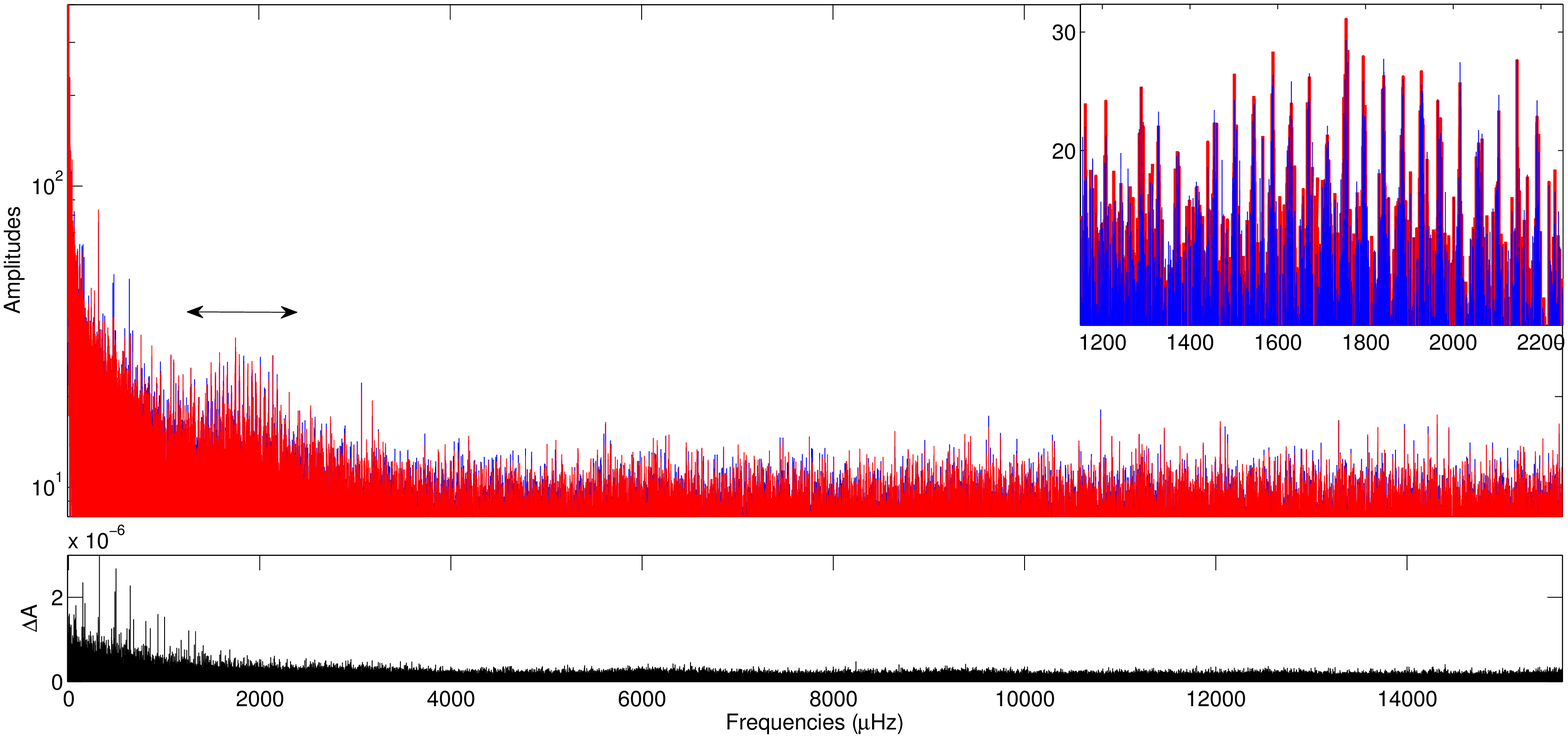}}
   \caption{Periodogram of \starc\ coming from linear interpolation (blue), and \arma\ interpolation (red). Notice that the order is inverted in the inset because the peaks in the p-mode region are higher for the periodogram of \arma\ interpolated data.}
   \label{starc-per}
\end{figure*}

The comparison of power spectra obtained with the two interpolation methods (Fig.~\ref{starc-per}) looks quite similar with slightly higher amplitudes for the ARMA interpolation (see inset of the figure) . However, a detail study of their difference using the absolute value of the amplitudes ($\Delta A= |A(ARMA) - A(LIN)|$) shows negligible non-systematic differences (Fig.~\ref{starc-per}, lower panel).


\section{Discussion}
Using a method of interpolation capable to preserve the original frequency content like MIARMA could be determining in order to perform seismic studies of A-F stars as, for example, the determination of periodicities in the pulsational spectrum of \ds\ stars \citep{AGH09}. Indeed, it has been shown \citep{JPG14a} that the amplitudes of the periodicities between frequencies are about 20$\%$ lower when linear interpolation is used for filling the gaps instead \arma\ interpolation.


We have shown that linear interpolation is insufficient to eliminate the effects of the spectral window in the power spectrum of classical pulsating stars observed by \corot. On the other side, the effects of the linear interpolation on the spectral window are insignificant in the case of solar-like stars. Nevertheless, the contribution of the signal found by \arma\ is not included when a linear interpolation is made so changing the original frequency content. Definitively, in order to fulfill the necessary condition for asteroseismology, it is necessary to use a gap-filling that preserves the original frequency content of the signal whatever the range of the excited pulsational frequencies.

\begin{figure*}	
   \centering
   \resizebox{\hsize}{!}{\includegraphics{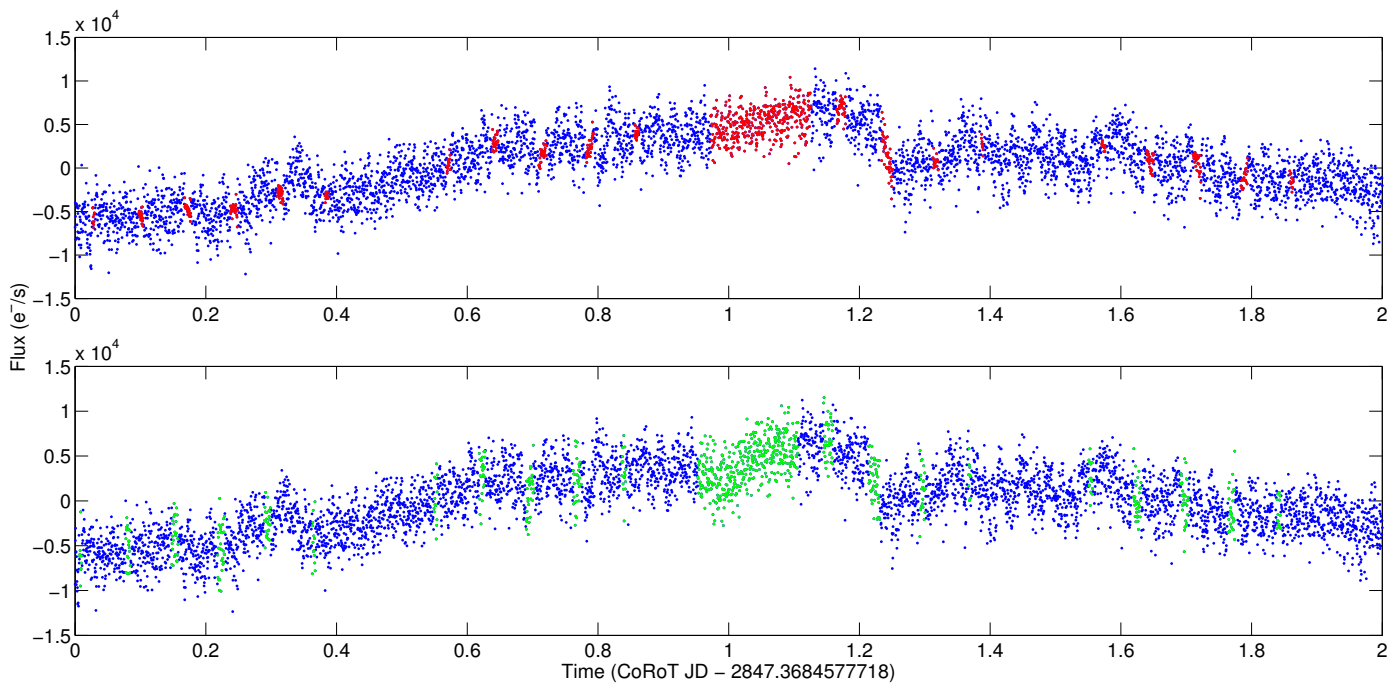}}
   \caption{Comparison between \arma\ interpolation (upper panel in red) and inpainting (lower panel in green) for the light curve of \starc\ observed by the \corot\ satellite. Valid data is depicted in blue (both panels).}    
   \label{starc-lc2}
\end{figure*}

The most recent reference to a gap-filling method in asteroseismology is the k-inpainting algorithm, presently used for \kepler\ data \citep{GAR14}. This algorithm is based on a sparsity prior of wavelets functions . We compared the results obtained by using this technique on the \corot\ data of \starc\ with those obtained by using MIARMA (see Fig.~\ref{starc-lc2}). Apparently the fine-scale structure observed in the light curve of \starc\ is preserved when performing both methods. Nevertheless, as the essential difference between the two approaches is that K-inpainting is based on analytic functions (namely Discrete Cosinus Transform) and MIARMA is not, we decided to test the ability of discerning noise from signal of both methods.

\begin{figure}	
   \centering
   \resizebox{\hsize}{!}{\includegraphics{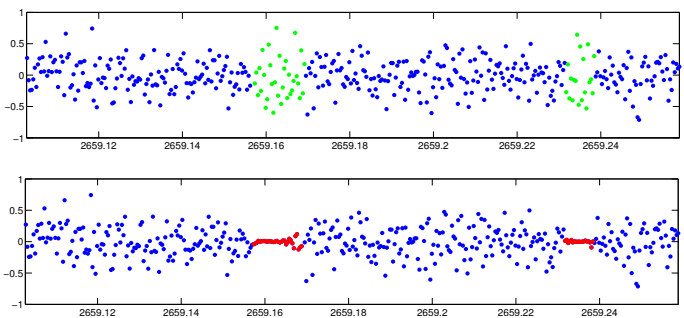}}
   \caption{Interpolation of gaps in a time series originated from a gaussian white noise process. Upper panel shows results from the inpainting technique and lower panel shows results from ARMA interpolation.}    
   \label{test}
\end{figure}

Both methods are used to interpolate gaps introduced in a purely random normally distributed time series (see Fig.~\ref{test}) with $\sigma = 0,2381$. As the necessary condition for gap-filling is to preserve the original frequency content of the signal, a proper interpolation should be zero. This only happens with \arma\ interpolation ($\sigma = 0,0753$, the small deviations from zero due to numerical error). On the other hand, the inpainting algorithm tries to mimick the completely random variations of the data ($\sigma = 0,2730$) thus failing to preserve the signal. This test demonstrates that the k-inpainting algorithm changes the original content of the signal.

On the light of this test the interpolation obtained with the inpainting technique in Fig.~\ref{starc-lc2} is not reliable because it is not clear whether it is reproducing the signal or the intrinsic noise of the time series. At the same time, Akaike criterion proves to be relevant to distinguish between white noise and signal as has been shown by the test. Hence, clearly the fine-scale structure interpolated by using \arma\ is not a white noise process. In any case these are novel features with unknown origin. We can speculate that the origin could be connected to the non-analyticity revealed in previous work of the authors \citep{JPG14b} in A-F stars. This is something that has to be examined more deeply in future studies.


\section{Conclusions}
Small and periodic gaps in time series of pulsating stars released by the space photometric mission \corot\ produce aliases of the satellite orbital period in the power spectra. Due to the very high duty cycle of the instrument, the amplitudes of these aliases are very small.
Linear interpolation is commonly used for filling those gaps in order to remove the aliasing effect. Here we demonstrate that while this approach works for high frequency pulsators (e.g. solar-like stars), it is not the case for longer period variables (e.g. A-F, Be stars, etc.), for which the aliases remain in the power spectra.

We present here MIARMA, a new method for filling gaps based on ARMA processes which preserve the original information contained 
in the time series. When the gaps of the CoRoT light curves are filled using this algorithm, all the aliases are eliminated whatever the type of variable considered. This is the reason why the method has been accepted by the \corot\ Scientific Committee for its implementation in the next correction of the data gathered.

When MIARMA interpolates data corresponding to the solar-like \starc\ the algorithm recognise as a signal the fine structure seen at a very short time scales. Other algorithms like inpainting provides similar results. However when a gapped time series of pure white noise is analysed, inpainting interpolates white noise datapoints while ARMA gives the correct answer, namely zeros.

%
%

\begin{acknowledgements}
      The authors acknowledge support from the "Plan Nacional de Investigación" under project AYA2012-39346-C02-01, and from the "Junta de Andaluc\'{i}a" local government under project 2012-P12-TIC-2469.
\end{acknowledgements}

\bibliographystyle{aa}
\bibliography{interpol}

\end{document}